\documentclass[intlimits,twoside,a4paper]{article}

\usepackage{amsmath,amssymb}
\usepackage{graphicx}

\usepackage[T2A]{fontenc}
\usepackage[cp1251]{inputenc}

\usepackage[eqsecnum]{cmpj3}

\issue{2017}{20}{4}{43301}
\doinumber{10.5488/CMP.20.43301}

\title[Interaction potential in methanol]%
{Modelling of potentials for interparticle interactions between methanol molecules
	
}

\author[N.P. Malomuzh, M.V. Timofeev]{N.P. Malomuzh, M.V. Timofeev}
\address{Odesa I.I. Mechnikov National University, 2 Dvoryans’ka St., 65026 Odesa, Ukraine}

\date{Received April 5, 2017, in final form August 20, 2017}

\begin{document}
	
\maketitle
	
\begin{abstract}
Peculiarities of interparticle interactions between methanol molecules in the methanol vapor are investigated. The bare potential is considered as a sum of repulsive, dispersive and electrostatic forces. It is supposed that H-bond is of electrostatic nature (the irreducible contribution caused by overlapping of electronic shells is unessential). The dispersive interaction is approximated with London’s formula, the electrostatic interaction is modelled by a multipole expansion up to dipole-octupole contribution. The multipole moments are assumed to be equal to their experimental values or to quantum chemical calculations. The repulsion is modelled by power potential, whose parameters  are fitted to the second virial coefficient and to the parameters of dimers. Along with the bare potential, the averaged potential of interparticle interaction is analyzed. It is shown that the repulsive potential has an exponent $n=28$. The multipole potential, presented in this paper, is scrupulously compared with the potential known as the OPLS.


\pacs 34.20.Gj, 51.30.+i
\end{abstract}

\section{Introduction}

Construction of a consistent statistical theory of methanol in liquid and vapor phase and water-alcohol solutions is connected with the problem of correct determination of interparticle interaction potentials. Molecules of water and alcohols are polyatomic. Therefore, their interparticle potentials depend on the distance between molecules and their relative orientations.

Various potentials are proposed for the description of interparticle interactions in methanol: OPLS~\cite{OPLS}, TIP \cite{TIP}, PHH3 \cite{PHH3}, H1 and H2~\cite{Haughney86, Haughney87}, coarse-grained model~\cite{Hus2014:PRE, Hus2014:JCP, Hus2015} and others. It is necessary to note that the potential OPLS allows one to successfully reproduce the second virial coefficient. However, the potentials~\cite{OPLS, TIP, PHH3, Haughney86, Haughney87} are not quite satisfactory since multipole moments corresponding to their effective charges essentially differ from the ones determined experimentally or with the help of quantum chemistry. Unfortunately, relevant parameters of these potentials are fitted in order to reproduce some properties of a liquid methanol. The potential proposed in~\cite{Hus2014:PRE, Hus2014:JCP, Hus2015} also allows one to reproduce the properties of aqueous solutions of methanol. However, the physical interpretation of different contributions therein is not always clear. The same is related to the analytical structure of the potentials. Namely, this circumstance is one of the reasons urging us to construct more suitable potentials.

Though the main attention in literature is paid to bare potentials, we should note that the thermodynamic properties of gases and liquids are mainly determined by the averaged potentials. Minor corrections to the thermodynamic variables are also caused by angular correlations~\cite{Lishchuk2010}. This  is due to continuous thermal rotation of molecules, whose characteristic rotation time  is noticeably smaller than that for the mean free time. Indeed, the characteristic time of the rotation of molecules is $\tau_{\text r} \sim \frac{2\piup }{\omega_{\text r} } \sim 2\piup (I/k_{\textrm{B}} T )^{1/2} $, where $I$ is the moment of inertia of a molecule. Mean free time is estimated according to $\tau_{\text f} \sim 1/(\sqrt{2} \piup d^2 n\bar{\upsilon }) $, where $\sigma$ is the effective diameter of a molecule, $n$ is the number density of molecules, $\bar{\upsilon }=\sqrt{3k_{\textrm{B}} T/m }$ is the average velocity of their motion. Inequality $\tau_{\text r} \ll\tau_{\text f}$ holds up to the density $n\ll n_*$ where $n_* \sim 2\cdot 10^{21}$~cm$^{-3}$ for methanol. The limiting value of the density  $n_*$  approaches the density in the triple point $n^{\text{(m)}} \sim 1.7\cdot 10^{22}$~cm$^{-3}$. This testifies to the validity of the ratio $\tau_{\text r} \ll \tau_{\text f}$  practically within the whole interval of vapor states from the triple point to the critical point.

The averaged potential $U(r)$ of intermolecular interaction is defined by the expression
\begin{equation}
\label{eq-averaged-pot}
\oint_{\Omega_{1} =4\piup }\frac{\rd\Omega_{1} }{4\piup }  \oint_{\Omega_{2} =4\piup }\frac{\rd\Omega_{2} }{4\piup }  \exp \left[-\beta \Phi \left(r,\Omega_{1} ,\Omega_{2} \right)\right]=\exp \left[-\beta U\left(r\right)\right],
\end{equation}
where $\Phi \left(r,\Omega_{1} ,\Omega_{2} \right)$ is the bare potential depending on angular variables $\Omega_{1}$ and $\Omega_{2}$, $r$ is the distance between the centers of mass of molecules,  $\beta =1/k_{\textrm{B}} T$. Such a definition of the averaged potential is genetically connected with the physical requirement: the free energy and the configurational integral corresponding to it should not depend on the choice of bare or averaged potentials in the pair approximation.

It should be stressed that the analysis of thermodynamic properties of liquids, in particular in methanol and aqueous solutions of methanol, should be based on the properties of the averaged potentials. In this case, we are able to make use of the similarity principle allowing us to establish the related properties of different alcohols and their aqueous solutions.

The aim of the present work is: 1) to construct a simple bare potential for methanol, all terms  of which are of clear physical nature. We assume that such a potential can be identified with a multipole expansion including the multipole moments determined experimentally or by using  quantum chemistry methods; 2) to construct  the averaged interaction potential; 3) to reproduce  the temperature dependence of the second virial coefficient and  properties of dimers as well as the critical temperature.

\section{Bare potential of the interparticle interaction in methanol}

The potential of interparticle interaction $\Phi$ in methanol vapor has the following structure
\begin{equation}
\label{eq-bare-pot}
\Phi =\Phi_{\textrm{R}} +\Phi_{\textrm{D}} +\Phi_{\textrm{E}}\,,
\end{equation}
where $\Phi_{\textrm{R}}$ is the repulsive potential, $\Phi_{\textrm{D}}$ is the term describing the dispersive interactions, $\Phi_{\textrm{E}}$ is the energy of the electrostatic interaction. H-bonds are considered as a sum of electrostatic interaction and an irreducible one caused by exchange effects arising due to the overlapping of the electronic shells of molecules.

It was shown in many works~\cite{Sokolov55, Dolgushin77, Fulton98, Barnes78, Hcontr, WaterPot} that H-bond in water is mainly of electrostatic nature. The irreducible part of H-bond, caused by the overlapping of electron shells and corresponding exchange effects, does not exceed $10{-}15$\% \cite{Hcontr}. Its contribution to thermodynamic potentials can be taken into account by the perturbation theory. The physical nature of H-bonds in methanol is close to that in water (the characteristic distances between oxygens and hydrogens are practically the same). Therefore, we conclude that the irreducible non-electrostatic part of H-bonding in methanol and aqueous solutions of methanol should be relatively small.

To describe the electrostatic interaction between methanol molecules, we use the multipole expansion. This potential will be referred to as the multipole potential (MP).

\vspace{\baselineskip}
\textit{a) Repulsion and dispersive interaction}
\vspace{\baselineskip}

The dispersive interaction will be modelled by the expression
\begin{equation}
\label{eq-disp-pot}
\Phi_{\textrm{D}} (r)=-\frac{A_{\textrm{D}}}{r^6}\,,
\end{equation}
in which $r$ is the distance between the centers of mass of methanol molecules, the value of the coefficient $A_{\textrm{D}}$ is estimated by London's approximation $A_{\textrm{D}} = 3/4 I\alpha^2$, where $I$ is the ionization potential, $\alpha$ is the electronic polarizability. For methanol molecule $\alpha = 3.2$~{\AA}$^3$ \cite{Nicolskii}, $I=10.85$~eV \cite{Nicolskii}, $\tilde{A}_{\textrm{D}} = 2.864$. Here, $\tilde{A}_{\textrm{D}} =A_{\textrm{D}} /k_{\textrm{B}} T_{\textrm{c}} [r_{\textrm{OO}}^{(\textrm{m})} ]^{6} $, $T_{\textrm{c}} = 512.6$ K is the critical temperature of methanol \cite{metCritT}, $r_{\textrm{OO}}^{(\textrm{m})} = 2.95$~{\AA} is the distance between oxygens in methanol dimer according to the calculations in 6-31G* basis \cite{MetDimer}.

The potential of the intermolecular repulsion is modelled by the power potential
\begin{equation}
\label{eq-rep-pot}
\Phi_{\textrm{R}} (r)=\frac{A_{\textrm{R}}}{r^n}\,.
\end{equation}
The parameters of this potential are chosen from the following conditions: 1) reproduction of the temperature dependence of the second virial coefficient, and 2) reproduction of the dimer equilibrium parameters of methanol. Here, $r$ is the distance between the centers of mass of the methanol molecules. A reduced value of $A_{\textrm{R}}$ is defined by the expression $\tilde{A}_{\textrm{R}} (n)=A_{\textrm{R}} /k_{\textrm{B}} T_{\textrm{c}} [r_{\text{OO}}^{(\textrm{m})} ]^{n} $, reduced distance: $\tilde{r}=r/r_{\textrm{OO}}^{(\textrm{m})} $.

\vspace{\baselineskip}
\textit{b) Modelling electrostatic interaction by the multipole expansion}
\vspace{\baselineskip}

\begin{figure}[!t]
	\centerline{\includegraphics[width=0.33\textwidth]{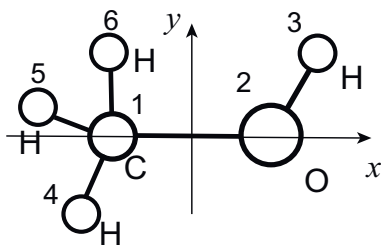}}
	\caption{Axes directions of MCS for methanol molecule (all atoms, except 5 and 6, are located in the plane of the picture).}
	\label{fig-molecule}
\end{figure}

In the multipole expansion of the electrostatic interaction energy, we will consider all contributions up to a dipole-octupole interaction $\Phi_{\textrm{DO}}$:
\begin{equation}
\label{eq-mult-pot}
\Phi_{\textrm{E}} =\Phi_{\textrm{DD}} +\Phi_{\textrm{DQ}} +\Phi_{\textrm{QQ}} +\Phi_{\textrm{DO}}\,.
\end{equation}
The explicit form of these terms is shown in \cite{WaterPot}. Multipole moments, that correspond to the location of the effective charges in such potentials as TIP \cite{TIP}, OPLS \cite{OPLS}, and others are much different from those obtained experimentally or by using quantum chemistry methods (see tables~\ref{tab-met-dip}--\ref{tab-met-oct}). In our calculations, we will use the multipole moments from \cite{MetDipol} and \cite{MetMultip}. The components of multipole moments are calculated in the molecular coordinate system (MCS), shown in figure~\ref{fig-molecule}. The origin of MSC is located at the center of mass (CM) of the molecule, and the mutual arrangement of atoms corresponds to \cite{MetGeom}.
\begin{table}[!b]
	\caption{Components of the dipole moment of an isolated molecule of methanol.}
	\label{tab-met-dip}
	\begin{center}
		\renewcommand{\arraystretch}{0}
		\begin{tabular}{|c|c|c|c|c|}
			\hline
			\hline
			& $d$, D & $d_x$, D & $d_y$, D & $d_z$, D \strut\\
			\hline
			\rule{0pt}{2pt}&&&&\\
			\hline
			Shtark effect, \cite{MetDipol} & 1.69 & $-0.885$ & 1.44 & 0 \strut\\
			\hline
			OPLS, \cite{OPLS} & 2.22 & $-1.19$ & 1.87 & 0 \strut\\
			\hline
			\hline
		\end{tabular}
		\renewcommand{\arraystretch}{1}
	\end{center}
\end{table}
\begin{table}[!b]
	\caption{Components of the quadrupole moment of an isolated molecule of methanol.}
	\label{tab-met-quadr}
	\begin{center}
		\renewcommand{\arraystretch}{1.2}
		\begin{tabular}{|c|c|c|c|c|}
			\hline
			\hline
			& $Q_{xx}$, $\text{D}\cdot\text{\AA}$ & $Q_{yy}$, $\text{D}\cdot\text{\AA}$ & $Q_{zz}$, $\text{D}\cdot\text{\AA}$ & $Q_{xy}$, $\text{D}\cdot\text{\AA}$ \\
			\hline
			\hline
			Quant. chem., \cite{MetMultip} & $-0.2393$ & 1.5419 & $-1.3026$ & 3.1487 \strut\\
			\hline
			OPLS, \cite{OPLS} & 0.426 & 0.965 & $-1.391$ & 2.745 \strut\\
			\hline
			\hline
		\end{tabular}
		\renewcommand{\arraystretch}{1}
	\end{center}
\end{table}
\begin{table}[!b]
	\caption{Components of the octupole moment of the isolated molecule of methanol.}
	\label{tab-met-oct}
	\begin{center}
		\begin{tabular}{|c|c|c|c|c|c|c|}
			\hline
			\hline
			& $O_{xxx}$,  & $O_{xyy}$, & $O_{xzz}$,  & $O_{xxy}$,  & $O_{xyy}$,  & $O_{xyz}$,  \\ 
			&$\text{D}\cdot\text{\AA}^2$&$\text{D}\cdot\text{\AA}^2$&$\text{D}\cdot\text{\AA}^2$&$\text{D}\cdot\text{\AA}^2$&$\text{D}\cdot\text{\AA}^2$&$\text{D}\cdot\text{\AA}^2$\\ 
			\hline
			\hline
			Quant. chem., \cite{MetMultip} & $-0.152$ & 2.8 & $-2.648$ & 3.214 & $-2.765$ & $-0.449$  \strut\\
			\hline
			\hline
		\end{tabular}
		\renewcommand{\arraystretch}{1}
	\end{center}
\end{table}

Note that close to the minimum, the behavior of a bare interparticle potential  cannot be described in the framework of model potentials including a small number of point-like charges or multipole moments. Therefore, the depth of a potential well cannot be determined with the help of model potentials. For this purpose, we make use of the ground state energy and dipole moment of a dimer. In this connection  we note that the multipole expansion quite satisfactorily approximate the electrostatic interaction between molecules up to distances $r > 1.1 r_{\text{min}}$\,,  where $r_{\text{min}}$ is the minimum position. For these distances, the energies of multipole contributions to the total potential satisfy the inequalities: ${|\Phi_{\text{DO}}|~<(\ll)~|\Phi_{\text{QQ}}|~<(\ll)~|\Phi_{\text{DQ}}|~<(\ll)~|\Phi_{\text{DD}}|}$. This is the condition for the applicability of multipole expansion. Some additional details can be found in \cite{Hcontr} and \cite{MultUsr}.

\vspace{\baselineskip}
\textit{c) Description of the methanol dimer and the second virial coefficient of methanol vapor}
\vspace{\baselineskip}

\begin{figure}[!b]
	\centerline{\includegraphics[width=0.33\textwidth]{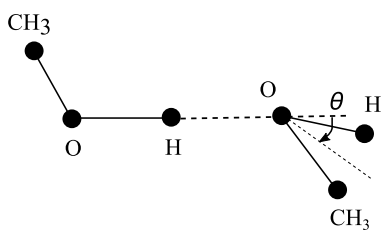}}
	\caption{ The relative position of molecules in the methanol dimer.}
	\label{fig-dimer}
\end{figure}

As it follows from our consideration, the parameters $A_{\textrm{R}}$, $n$ remain undetermined. To determine these values we will use the energy of the ground state of the dimer, its dipole moment and the temperature dependence of the second virial coefficient. The characteristics of the dimer from different sources are presented in table~\ref{tab-met-dimer}.
The relative positioning of methanol molecules corresponding to the considered dimer configuration is presented in figure~\ref{fig-dimer}.

The values of $A_{\textrm{R}}$ for different $n$ are determined by the least squares method for reproduction of temperature dependence of the second virial coefficient. The obtained values of $A_{\textrm{R}}$ at various exponents~$n$, as well as the main characteristics of the dimer, are placed in table~\ref{tab-met-dimer2}. Here, $\tilde{r}_{\textrm{c.m.}} = r_{\textrm{c.m.}} /r_{\textrm{OO}}^{(\textrm{m})} $ is the distance between the centers of mass of molecules, $\tilde{r}_{\textrm{OO}} = r_{\textrm{OO}} /r_{\textrm{OO}}^{(\textrm{m})} $ is the distance between oxygens, $\tilde{E} = E / k_{\textrm{B}} T_{\textrm{c}}$ is the energy of the dimer ground state.

\renewcommand{\arraystretch}{1.2}
\begin{table}[!t]
	\caption{A comparative characteristic of the methanol dimer.}
	\label{tab-met-dimer}
	\begin{center}
		\begin{tabular}{|c|c|c|c|c|c|}
			\hline\hline
			& $\tilde{r}_{\textrm{OO}} $ & $\theta ,\, ^{\circ } $ & $\tilde{E}$ & $d$, D & $d_{\textrm{D}}$, D \\
			\hline
			\hline
			STO-3G (see \cite{MetDimer}) & 0.929 & 48 & $-5.47$ & 1.51 & 2.9 \strut\\
			\hline
			4-31G  (see \cite{MetDimer}) & 0.956 & 28 & $-7.91$ & 2.36 & 4.16 \strut\\
			\hline
			6-31G* (see \cite{MetDimer}) & 1.0 & 48 & $-5.56$ & 1.94 & 3.02 \strut\\
			\hline
			OPLS \cite{OPLS} & 0.925 & 22 & $-6.67$ & 2.22 & 3.6 \strut\\
			\hline
			MP & 0.903 & 10.5 & $-5.85$ & 1.69 & 2.92 \strut\\
			\hline
			\hline
		\end{tabular}
			\end{center}
\end{table}
\renewcommand{\arraystretch}{1.2}
\begin{table}[!t]
	\caption{Characteristics of the methanol dimer at various exponents $n$.}
	\label{tab-met-dimer2}
	\begin{center}
		\begin{tabular}{|c|c|c|c|c|}
			\hline
			\hline
			$n$ &  12 & 18 &  24 & 28 \strut\\
			\hline
			\hline
			$\strut\tilde{A}_{\textrm{R}} \left(n\right)$ & 8.78 & 16.1 & 34.56 & 59.46 \strut\\
			\hline
			$\tilde{r}_{\textrm{OO}} $ & 0.823 & 0.875 & 0.895 & 0.902 \strut\\
			\hline
			$\tilde{r}_{\textrm{c.m.}} $ & 1.078 & 1.122 & 1.146 & 1.153 \strut\\
			\hline
			$\theta ,\, ^{\circ } $ & 10.2 & 10.4 & 10.5 & 10.5 \strut\\
			\hline
			$\tilde{E}$ & 5.87 & 5.84 & 5.83 & 5.85 \strut\\
			\hline
			$d_{\textrm{D}}$, D & 2.92 & 2.92 & 2.92 & 2.92 \strut\\
			\hline
			\hline
		\end{tabular}
		\renewcommand{\arraystretch}{1}
	\end{center}
\end{table}

The best consent of the obtained distance between oxygens of methanol molecules with value of 6-31G* is reached at exponent $n=28$ in the repulsion potential (\ref{eq-rep-pot}).

\section{Averaged interaction potential of molecules of a methanol}

In this section, the numerical values of the averaged potential received by equation (\ref{eq-averaged-pot}) will be approximated with the analytical expression. First of all, it is considered that the distant tail of the potential is caused by the dipole-dipole interaction. At smaller distances, the terms caused by a dipole-quadrupole and a quadrupole-quadrupole interaction will also be considered. According to the above, the potential of the averaged intermolecular interaction is approximated by the expression:
\begin{equation}
\label{eq-appr-pot}
U(r)=\alpha \left[\alpha_{28} \left(\frac{\sigma }{r} \right)^{28} -\left(\frac{\sigma }{r} \right)^{6} -\alpha_{8} \left(\frac{\sigma }{r} \right)^{8} -\alpha_{10} \left(\frac{\sigma }{r} \right)^{10} \right].
\end{equation}
The value of the combination $\alpha \alpha_{28} \sigma^{28} / k_{\textrm{B}} T_{\textrm{c}} [r_\textrm{OO}^{(\textrm{m})}]^{28} = 59.46$ is the same as for the bare potential since the repulsion is described by the contribution which does not depend on angular variables. The distant asymptotic of the attraction potential corresponds to the dispersion and the dipole-dipole interactions. The value of the combination $\alpha \sigma^{6}$ is calculated for the distant tail of the averaged potential  (\ref{eq-averaged-pot}) and the values of $\alpha$ are placed in table~\ref{tab-aver-pars}. The values of coefficients $\alpha_{8}$, $\alpha _{10}$, are selected so that 1) they satisfy a ratio $1 > \alpha_{8} > \alpha_{10} > 0$ and 2) they approximate the behavior of the averaged potential. The specified inequality ensures the convergence of the asymptotic multipolar series for the averaged potential. The obtained values of coefficients are shown in table~\ref{tab-aver-pars}. The values of parameter $\sigma$ correspond to the distances at which the averaged potential equals zero. Here, $\tilde{T} = T / T_{\textrm{c}}$.

\begin{figure}[!t]
	\centerline{\includegraphics[width=0.5\textwidth]{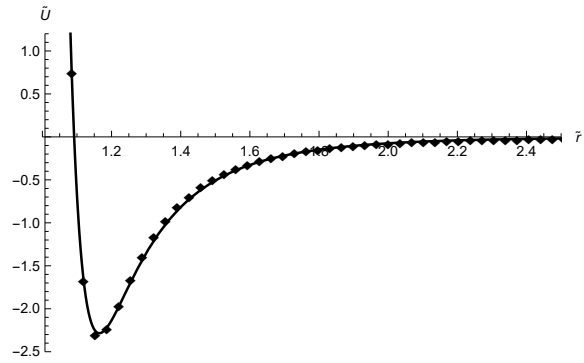}}
	\caption{Averaged potential of the intermolecular interaction at $\tilde{T} = 0.63$. Points are numerical values, the curve is their approximation.}
	\label{fig-averaged}
\end{figure}

The approximation of the averaged potential by expression (\ref{eq-appr-pot}) is shown in figure~\ref{fig-averaged}.

\renewcommand{\arraystretch}{1.2}
\begin{table}[!h]
	\caption{Values of parameters for (\ref{eq-appr-pot}) at various $\tilde{T}$.}
	\label{tab-aver-pars}
	\begin{center}
		\begin{tabular}{|c|c|c|c|c|c|c|}
			\hline
			\hline
			$\tilde{T}$ & $\tilde{\sigma}$ & $\alpha$ & $\alpha_{28} $ & $\alpha_{8} $ & $\alpha_{10} $ & $\tilde{U}_{\textrm{min}}$ \strut\\
			\hline
			\hline
			0.63 & 1.088 & 4.52 & 2.35 & 0.692 & 0.686 & $-2.57$ \strut\\
			\hline
			0.728 & 1.092 & 4.29 & 2.25 & 0.616 & 0.616 & $-2.28$ \strut\\
			\hline
			0.825 & 1.096 & 4.08 & 2.123 & 0.548 & 0.548 & $-2.06$ \strut\\
			\hline
			0.923 & 1.099 & 3.91 & 2.007 & 0.492 & 0.492 & $-1.87$ \strut\\
			\hline
			\hline
		\end{tabular}
		\renewcommand{\arraystretch}{1}
	\end{center}
\end{table}

\section{Comparative characteristics of the MP and OPLS}

Let us compare the properties of the MP (\ref{eq-bare-pot}) with those for the OPLS and quantum chemical calculations. For this purpose, we consider the characteristics of an isolated dimer, asymptotical behavior of the averaged potential, temperature dependence of the second virial coefficient and estimates for the critical temperature of methanol.

This comparison cannot be quite correct since parameters of the OPLS are fitted to reproduce some properties of the liquid phase. In our case, the interaction of isolated methanol molecules is considered.

\subsection{Dimer characteristics}

To  the most important characteristics of a dimer we relate: dipole moment of an isolated methanol molecule $d$, dipole moment of methanol dimer $d_{\textrm{D}}$,  dimer ground state energy $E$, the distance between oxygens $r_{\textrm{OO}}$. Their numerical values obtained with the help of the MP and OPLS as well as calculated by quantum chemical method in \cite{MetDimer} are presented in table \ref{tab-met-dimer}. Note that different modifications of the Hartree-Fock method (different bases) are used in \cite{MetDimer}.

The value of the dipole moment and the ground state energy are the most important characteristics of a dimer. Thereto, they both are better reproduced  with the help of the MP. Indeed, the dipole moment of an isolated molecule, following from the OPLS, exceeds its experimental value ($d=1.69$~D) by more than $30\%$. Similarly, the dimer dipole moment more than two times exceeds the one for an isolated molecule. At the same time, the dipole moment of an isolated molecule within the MP coincides with its experimental value and the determined value of the dimer dipole moment is close to that  obtained with the help of quantum-chemical calculations.

It is important that the so-called ``H-bond'', connecting two methanol molecules into dimer, is formed by the same groups of atoms as in water. Parameters of these H-bonds are also close. Therefore, we expect that the ground state energy for methanol dimer should be close to that for water dimer ($\tilde{E}=5.1{-}5.3$). This assumption is confirmed by the 6-31G* calculation. The MP leads to a close result while the OPLS yields a much greater value.

\subsection{Asymptotical behavior of averaged potentials}

Let us compare the averaged potentials of the electrostatic interaction in the multipole approximation. The averaged potential of the multipole interaction is determined by the expression \cite{MultUsr}:

\begin{equation}
\label{eq:mult-usr-pot}
\tilde{U}_{\textrm{el}} = -\frac{1}{\tilde{T}} \left( \frac{\tilde{A}_{6}}{\tilde{r}^{6}} + \frac{\tilde{A}_{8}}{\tilde{r}^{8}} +  \frac{\tilde{A}_{10}}{\tilde{r}^{10}} \right),
\end{equation}
where $\tilde{A}_{6} = \frac{2}{3} \frac{(d^{2})^{2}}{\sigma_{r}^{6} (k_{\textrm{B}} T_{\textrm{c}})^{2}}$,  $\tilde{A}_{8} = 3 \frac{d^{2} Q_{\alpha\beta}^{2}}{\sigma_{r}^{6} (k_{\textrm{B}} T_{\textrm{c}})^{2}}$, $\tilde{A}_{10} = \frac{63}{10} \frac{(Q_{\alpha\beta}^{2})^{2}}{\sigma_{r}^{6} (k_{\textrm{B}} T_{\textrm{c}})^{2}}$, $Q_{\alpha \beta }^2 = Q_{\alpha \beta } Q_{\beta \alpha }$, $Q_{\alpha \beta }$ are components of the quadrupole moment tensor in the molecular coordinate system, whose center is located at the center of mass of the methanol molecule, $\sigma_{r}=r_{\textrm{OO}}^{(\textrm{m})}$, $r$ is the distance between the centers of mass of the molecules of methanol. The components of the quadrupole moment for the MP and OPLS are shown in  table \ref{tab-met-quadr}. The values of the dipole moments are given in  table \ref{tab-met-dip}.
			
For the MP and OPLS, these coefficients presented in the table \ref{tab-usr-mult-coef}.

\begin{table}[!t]
\begin{center}
	\caption{Values of $\tilde{A}_{6}$, $\tilde{A}_{8}$, $\tilde{A}_{10}$ for (\ref{eq:mult-usr-pot}).}
	\label{tab-usr-mult-coef}
	\vspace{2ex}
		\renewcommand{\arraystretch}{1.2}
		\begin{tabular}{|c|c|c|c|}
			\hline
			\hline
			& $\tilde{A}_{6}$ & $\tilde{A}_{8}$ & $\tilde{A}_{10}$ \strut\\
			\hline
			\hline
			MP & 0.5 & 1.45 & 1.96 \strut\\
			\hline
			OPLS & 1.49 & 1.9 & 1.13 \strut\\
			\hline
			\hline
		\end{tabular}
		\renewcommand{\arraystretch}{1}
	\end{center}
\end{table}

Asymptotical behavior of the averaged potentials is determined by the dipole moments. However, corresponding contributions of the OPLS differ much from those of MP.

\subsection{Temperature dependence of the second virial coefficient}

In general, the second virial coefficient depends on the bare potential \cite{VirCoefExpr1}: 
\[B(T)=\frac{1}{2V\Omega _{0}^{2} } \int _{\Gamma }\left\{1-{\rm exp}\left[-\beta \Phi \left(r,\Omega_{1} ,\Omega_{2} \right)\right]\right\} \rd\Gamma _{1} \rd\Gamma_{2}\,, \]
where $\rd\Gamma_{i} = \rd\vec{r}_{i} \rd\Omega_{i}$, $\vec{r}_{i}$ is the position vector to $i$-th molecule, $\Omega_{0} $ is the phase volume corresponding to the particular choice of angular variables.

\begin{figure}[!h]
	\centerline{\includegraphics[width=0.5\textwidth]{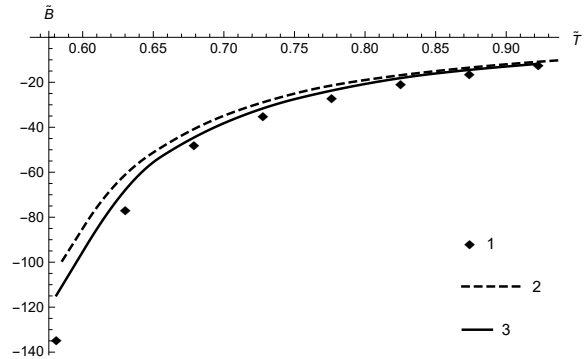}}
	\caption{Second virial coefficient of methanol vapor. 1 are experimental data \cite{MetVirCoef}, 2 is the second virial coefficient for OPLS potential \cite{OPLS-B}, 3 --- this work.}
	\label{fig-vir}
\end{figure}

It is easy to see that, after integration over the angular variables, the expression for the second virial coefficient is reduced to the standard form \cite{VirCoefExpr2}:
\begin{equation}
\label{vir-coef}
B(T)=2\piup \int_{r}\left[1-\re^{-\beta U \left(r\right)} \right]r^{2} \rd r,
\end{equation}
where $U$ is the averaged potential. The temperature dependencies of the second virial coefficient calculated with the help of the averaged potential (\ref{eq-appr-pot}) corresponding to the MP, data \cite{OPLS-B} for OPLS and experimental data \cite{MetVirCoef} are presented in figure \ref{fig-vir}. Here, the dimensionless value of $B$ is: $B / [r_{\textrm{OO}}^{(\textrm{m})}]^3$.

\subsection{The critical temperature of methanol}

In order to estimate the critical temperature of methanol we turn to the similarity principle not requiring detailed cumbersome calculations.

From its standard formulation \cite{Fisher} it follows that the critical temperatures of two atomic liquid types of argon satisfy the equation:
\begin{equation}
\label{eq:simil-T}
T_{\textrm{c}}^{(2)} = \frac{\varepsilon_2}{\varepsilon_1} T_{\textrm{c}}^{(1)},
\end{equation}
where $ \varepsilon_i = 1, 2 $ are the parameters determining the depth of the potential well for the interparticle potential:

\begin{equation}
\label{eq:argon-like-pot}
U = \varepsilon \phi \left( r/ \sigma \right).
\end{equation}
Here, $\sigma$ is the diameter of a molecule. The application of (\ref{eq:simil-T}) to atomic liquids leads to a quite satisfactory agreement with experimental data.

However, the intermediate application of (\ref{eq:simil-T}) to water, methanol and other alcohols is impossible since their interparticle potentials have a more complex structure (see \cite{SPC, TIP5P, OPLS}). In this case, it is necessary to pay attention to the fact that thermodynamic properties of low-molecular liquids with non-spherical particles are determined by the averaged potentials having the structure (\ref{eq:argon-like-pot}) (see  details in \cite{MultUsr, WaterPot}). Thereto, the averaged potentials are not fully equivalent to (\ref{eq:argon-like-pot}) since their parameters $\varepsilon_{\text a}$  and $\sigma_{\text a}$ depend on temperature.

In this connection,  we suppose, that 1) the similarity principle can be applied to all low-molecular liquids for which the averaged potentials assume the structure (\ref{eq:argon-like-pot}); 2) the critical temperatures of such liquids satisfy the equation similar to (\ref{eq:simil-T}):

\begin{equation}
\label{eq:simil-T-T}
T_{\textrm{c}}^{(2)} = \frac{\bar{\varepsilon}_{\text a}^{(2)} \big( T_{\textrm{c}}^{(2)} \big) }{\bar{\varepsilon}_{\text a}^{(1)} \big( T_{\textrm{c}}^{(1)} \big) } T_{\textrm{c}}^{(1)}
\end{equation}
and taking into account the temperature dependencies of the interaction constants; 3)  these constants are connected with the values $ \varepsilon_{\text a}^{(i)}$, $i = 1, 2 $ by the relations:

\begin{equation}
\label{eq:eps-dzeta}
\bar{\varepsilon}_{\text a}^{(i)} = \varepsilon_{\text a}^{(i)} / \zeta_{i}\,,
\end{equation}
where $\zeta_{i}$ is the high frequency value of the dielectric permittivity, allowing one to describe the  screening effects in dense media.

Moreover, we  suppose that $\zeta_{i}$ can be estimated according to the expression:

\begin{equation}
\label{eq:dzeta-ro}
\frac{\zeta(\rho) - 1}{\zeta(\rho) + 2} = \frac{4 \piup}{3} \frac{\rho}{m} \alpha,
\end{equation}
where $\alpha$ is the electronic polarizability of a molecule, $\rho$ is the mass density and $m$ is the molecular mass. This formula is similar to the Clausius-Mossotti one, being applicable only to atomic liquids.  Thus, it is assumed that the values of $\zeta_{i}$ are equal to that at frequencies of rotational motion of molecules, i.e., it is mainly determined by distortion of electronic shells of molecules. This circumstance is typical of the systems with non-polar molecules.

\begin{table}[!b]
	\caption{The estimates for the critical temperature for methanol ($\tilde{T}_{\textrm{c}}$).}
	\label{tab:T-crit}
	\begin{center}
		\renewcommand{\arraystretch}{0}
		\begin{tabular}{|c|c|c|}
			\hline
			\hline
			& MP & OPLS  \strut\\
			\hline
			\hline
			Argon basis & 1.014 & 1.11  \strut\\
			\hline
			Water basis & 1.005 & 0.991 \strut\\
			\hline
			\hline
		\end{tabular}
		\renewcommand{\arraystretch}{1}
	\end{center}
\end{table}

Finally, we get the following formula for the critical temperatures:
\begin{equation}
\label{eq:simil-T-final}
T_{\textrm{c}}^{(2)} = \frac{\varepsilon_{\text a}^{(2)} \big( T_{\textrm{c}}^{(2)} \big) \big/ \zeta_{2} \big( T_{\textrm{c}}^{(2)}  \big)}{\varepsilon_{\text a}^{(1)} \big( T_{\textrm{c}}^{(1)} \big) \big/ \zeta_{1} \big( T_{\textrm{c}}^{(1)}  \big)} T_{\textrm{c}}^{(1)},
\end{equation}
where $ \varepsilon_{\text a}^{(i)}$, $i = 1, 2 $, are calculated for two isolated molecules and $\zeta_{i} ( T_{\textrm{c}}^{(i)} )$, $i = 1, 2$, depend indirectly on temperature: $\zeta_{i} ( T_{\textrm{c}}^{(i)} ) \equiv \zeta_{i} ( \rho_{i} ( T_{\textrm{c}}^{(i)} ) )$.

We use the formula (\ref{eq:simil-T-final}) to determine the critical temperature of methanol, choosing parameters of argon and water as basis ones (see table \ref{tab:T-crit}). Note that the values of density for methanol and basic liquid correspond to their coexistence curves.

The experimental value of the methanol critical temperature is: $\tilde{T}_{\textrm{c}}^{(\textrm{m})} = 1.0$, i.e., in both cases the MP leads to a better agreement with this value. Note, that the OPLS is an effective potential, so it is not clear how to precisely take  the screening effects into account.

\section{Conclusion}

In this paper, we have constructed the bare and averaged potentials of intermolecular interaction between methanol molecules. Rigorously speaking, these potentials are only correct for a vapor phase, where the screening effect can be ignored. In order to build them we have used an approach similar to that for the water potential \cite{WaterPot}. It is shown that the averaged potential can be approximated by the expression, whose structure is close to the one for the Lennard-Jones potential. Let us note that following the distinctions of the obtained potentials from those in literature \cite{OPLS,TIP,PHH3}, 1) the repulsive potential has a power form and  is characterized by the exponent $n = 28$ and 2) the attractive contribution includes a standard term, $~1/r^{6}$, and terms, $~1/r^{8}$ and $~1/r^{10}$,  caused by dipole-quadrupole and quadrupole-quadrupole interactions. Note that contribution with $n = 28$ is also inherent to argon \cite{argonRep1} and water \cite{WaterPot}. The weak temperature dependence of parameters of the averaged potential is an additional feature of this potential.
This averaged potential with good accuracy reproduces the second virial coefficient of methanol vapor. The bare potential includes an experimental dipole moment and contains multipole moments of higher order that are determined by quantum chemical methods.

The presented potential is used to retrieve the dimer characteristic and calculate the second virial coefficient. Main parameters of methanol dimer, the energy of the ground state and its dipole moment, are better reproduced than with the help of OPLS potential. The second virial coefficient is also better reproduced by MP than by OPLS.

Introducing the averaged potentials allows us to formulate the generalized similarity principle. Applying this principle we calculated  the critical temperature of methanol with high accuracy, supposing that argon and water are basic liquids with all known properties. We get that MP leads to a better agreement with experimental data than OPLS.

The position of the critical point for methanol was also considered in \cite{OPLS-T1, OPLS-T2, OPLS-T3}. Therein, the binodal position was established too. For this purpose, molecular dynamics, Monte-Carlo and other methods were used. However, the values of the critical temperatures, obtained with help of the generalized similarity principle, are in a better agreement with experimental data. Note that the positions of binodal and spinodal for methanol can be also found with the help of the generalized similarity principle applied to the ones of argon or water.

The presented approach can be also used to build intermolecular interaction potentials in ethanol and other alcohols, as well as to investigate the interactions between water and alcohol molecules. The obtained potentials allow one to calculate the enthalpy of mixing of water with alcohols, contraction of  water-alcohol mixtures (see \cite{contraction}) and other thermodynamic properties.

\section*{Acknowledgements}
The authors cordially thank Professor Leonid Bulavin for a detailed discussion of the results obtained, senior research assistant Vitaliy Bardic for consultations on the repulsive potential behavior and Professor Mykhailo Kozlovskii for useful advice.

\ukrainianpart

\title{Моделювання потенціалу міжчастинкової взаємодії в метанолі}
\author{М.П. Маломуж, М.В. Тимофєєв}
\address{Одеський національний університет ім. І. І. Мечникова, вул. Дворянська, 2, 65026 Одеса, Україна}

\makeukrtitle

\begin{abstract}
	\tolerance=3000%
Досліджується структура і явний вигляд потенціалу міжчастинкової взаємодії в парі метанолу. Вихідний потенціал розглядається як сума сил відштовхування, дисперсійної та електростатичної взаємодії. Припускається, що водневі зв’язки мають електростатичну природу (незвідний внесок, обумовлений перекриттям електронних оболонок є незначним). Дисперсійна взаємодія апроксимується у наближенні Лондона, електростатична взаємодія моделюється за допомогою мультипольного розкладу до диполь-октупольного внеску включно. Використовуються експериментальні значення мультипольних моментів або результати квантово-хімічних розрахунків. Відштовхування моделюється степеневим потенціалом, параметри якого вибираються для відтворення другого віріального коефіцієнту та характеристик димеру. Разом з вихідним потенціалом, досліджується усереднений потенціал міжчастинкової взаємодії. Показано, що потенціал відштовхування має показник степеня $n=28$. Представлений мультипольний потенціал порівнюється з потенціалом OPLS.
	\keywords метанол, потенціал взаємодії
	
\end{abstract}


\begin{thebibliography}{99}
	\bibitem{OPLS} Jorgensen W.L., J. Phys. Chem., 1986, \textbf{90}, 1276, \doi{10.1021/j100398a015}.
	\bibitem{TIP} Jorgensen W.L., J. Am. Chem. Soc., 1981, \textbf{103}, 341, \doi{10.1021/ja00392a017}.
	\bibitem{PHH3} Palinkas G., Hawlicka E., Heinzinger K., J. Phys. Chem., 1987, \textbf{91}, 4334, \doi{10.1021/j100300a026}.
	\bibitem{Haughney86} Haughney M., Ferrario M., McDonald I.R., Mol. Phys., 1986, \textbf{58}, 849–853, \doi{10.1080/00268978600101611}.
	\bibitem{Haughney87} Haughney M., Ferrario M., McDonald I.R., J. Phys. Chem., 1987, \textbf{91}, 4934–4940, \doi{10.1021/j100303a011}.
	\bibitem{Hus2014:PRE} Hu{\v s} M., Urbic T., Phys. Rev. E, 2014, \textbf{90}, 062306, \doi{10.1103/PhysRevE.90.062306}.
	\bibitem{Hus2014:JCP} Hu{\v s} M., Muna{\` o} G., Urbic T., J. Chem. Phys., 2014, \textbf{141}, 164505, \doi{10.1063/1.4899316}.
	\bibitem{Hus2015} Hu{\v s} M., {\v Z}akelj G., Urbi{\v c} T., Acta Chim. Slov., 2015, \textbf{62}, 524–530, \doi{10.17344/acsi.2015.1441}.
	\bibitem{Lishchuk2010} Lishchuk S.V., Malomuzh N.P., Makhlaichuk P.V., Phys. Lett. A, 2010, \textbf{374}, No. 19–20, 2084–2088, \\ \doi{10.1016/j.physleta.2010.02.070}.
	\bibitem{Sokolov55} Sokolov N., Usp. Fiz. Nauk, 1955, \textbf{57}, 205--278 (in Russian), \bibdoi{10.3367/UFNr.0057.195510d.0205}.
	\bibitem{Dolgushin77} Dolgushin M., Preprint of the Inst. Teor. Fiz., ITF-77-83, Kiev, 1977, (in Russian).
	\bibitem{Fulton98} Fulton R.L., Perhaes P., J. Phys. Chem. A, 1998, \textbf{102}, 9001--9020, \doi{10.1021/jp9821228}.
	\bibitem{Barnes78} Barnes P., In: Progress in Liquid Physics, Croxton C.A. (Ed.), Wiley, Chichester, 1978, 391--428.
	\bibitem{Hcontr} Makhlaichuk P.V., Malomuzh M.P., Zhyganiuk I.V., Ukr. J. Phys., 2012, \textbf{57}, 113.
	\bibitem{WaterPot} Timofeev M.V.,  Ukr. J. Phys., 2016, \textbf{61}, 893, \doi{10.15407/ujpe61.10.0893}.
	\bibitem{Nicolskii} Nikol’skii B.P. (Ed.), The Chemist’s Handbook, Vol. 1, Khimiya, Moscow, 1966, (in Russian).
	\bibitem{metCritT} Craven R.J.B., de~Reuck K.M., Int. J. Thermophys., 1986, \textbf{7}, 541, \doi{10.1007/BF00502388}.
	\bibitem{MetDimer} Tse Y.{-}C., Newton M.D., Allen L.C., Chem. Phys. Lett., 1980, \textbf{75}, 350, \doi{10.1016/0009-2614(80)80529-X}.
	\bibitem{MetDipol} Ivash E.V., Dennison D.M., J. Chem. Phys., 1953, \textbf{21}, 1804, \doi{10.1063/1.1698668}.
	\bibitem{MetMultip} Huiszoon C., Mol. Phys., 1986, \textbf{58}, No.~5, 865, \doi{10.1080/00268978600101641}.
	\bibitem{MetGeom} Lees R.M., Baher J.G., J. Chem. Phys., 1968, \textbf{48}, 5299, \doi{10.1063/1.1668221}.
	\bibitem{MultUsr} Makhlaichuk P.V., Candidate’s Dissertation in Mathematics and Physics, Odesa, 2013.
	\bibitem{VirCoefExpr1} Hirschfelder J.O., Curtiss C.F., Bird R.B., The Molecular Theory of Gases and Liquids, Wiley, New York, 1954.
	\bibitem{VirCoefExpr2} Landau L.D., Lifshitz E.M., Statistical Physics, Part 1, Pergamon Press, Oxford, 1980. 
	\bibitem{OPLS-B} Buck U., Schmidt B., J. Chem. Phys., 1993, \textbf{98}, 9410, \doi{10.1063/1.464373}.
	\bibitem{MetVirCoef} Kudchadke A.P., Eubank P.T., J. Chem. Eng. Data, 1970, \textbf{15}, No.~1, 7, \doi{10.1021/je60044a005}.
	\bibitem{Fisher} Fisher I.Z., Statistical Theory of Liquids, University of Chicago Press, Chicago, 1964. 
	\bibitem{SPC} Berendsen H.J.C., Postma J.P.M., van Gunsteren W.F., Hermans J., In: Intermolecular Forces, Pullman~B.~(Ed.), Reidel, Dordrecht, 1981, 331--342, \doi{10.1007/978-94-015-7658-1_21}.
	\bibitem{TIP5P} Mahoney M.W., Jorgensen W.L., J. Chem. Phys., 2000, \textbf{112}, 8910, \doi{10.1063/1.481505}.
	\bibitem{argonRep1} Bardic V.Yu., Malomuzh N.P., Sysoev V.M., J. Mol. Phys., 2005, \textbf{120}, 27, \doi{10.1016/j.molliq.2004.07.020}.
	\bibitem{OPLS-T1} Kettler M., Nezbeda I., Chialvo A.A., Cummings P.T., J. Phys. Chem. B, 2002, \textbf{106}, 7537, \doi{10.1021/jp020139r}.
	\bibitem{OPLS-T2} Gonzalez Salgado D., Vega C., J. Chem. Phys., 2010, \textbf{132}, 094505, \doi{10.1063/1.3328667}.
	\bibitem{OPLS-T3} Gonzalez-Salgado D., Vega C., J. Chem. Phys., 2016, \textbf{145}, 034508, \doi{10.1063/1.4958320}.
	\bibitem{contraction} Gotsul’skii V.Ya., Malomuzh N.P., Chechko V.E., Russ. J. Phys. Chem. A, 2013, \textbf{87}, 1638,\\ \doi{10.1134/S0036024413100087}.
\end{thebibliography}
\end{document}